\documentclass[12pt]{iopart}
\newcommand{\msr}{$\mu$SR}
\newcommand{\lfaof}{LaFeAsO$_{1-x}$F$_x$}
\newcommand{\cfcaf}{CaFe$_{1-x}$Co$_x$AsF}
\newcommand{\bkfa}{Ba$_{1-x}$K$_x$Fe$_2$As$_2$}
\usepackage{iopams}  
\usepackage{graphicx}
\begin{document}

\title[Magnetism and superconductivity in iron pnictides probed by \msr]{Competition/Coexisitence of Magnetism and Superconductivity in 
Iron Pnictides Probed by Muon Spin Rotation
}

\author{Soshi Takeshita, Ryosuke Kadono}
\address{
Institute of Materials Structure Science, High Energy Accelerator
Research Organization (KEK), 1-1 Oho, Tsukuba, Ibaraki 305-0801, Japan}
\ead{soshi@post.kek.jp, ryosuke.kadono@kek.jp}
\begin{abstract}
The presence of macroscopic phase separation into superconducting and 
magnetic phases in \lfaof\ and \cfcaf\ is demonstrated by muon spin rotation (\msr) measurement across their phase boundaries ($x=0.06$ for \lfaof\  and $x=0.075$--0.15 for \cfcaf).  In LaFeAsO$_{0.94}$F$_{0.06}$, both magnetism and superconductivity develop simultaneously below a common critical temperature, $T_{\rm m}\simeq T_{\rm c}\simeq18$ K, where the magnetism is characterized by strong randomness.  A similar, but more distinct segregation of these two phases is observed in \cfcaf, where the magnetic phase retains $T_{\rm m}$ as close to that of the parent compound ($T_{\rm c}\ll T_{\rm m}\simeq80$--120 K) and the superconducting volume fraction is mostly proportional to the Co content $x$.
The close relationship between magnetism and superconductivity is discussed based on these experimental observations. 
Concerning superconducting phase, an assessment is made on the anisotropy of order parameter in the superconducting state of \lfaof, \cfcaf, and \bkfa\ ($x=0.4$) based on the temperature dependence of superfluid density [$n_{\rm s}(T)$] measured by \msr. The gap parameter, 
$2\Delta/k_BT_c$, determined from $n_{\rm s}(T)$ exhibits a tendency that hole-doped pnictides (\bkfa) is much greater than those in electron-doped ones (\lfaof, \cfcaf), suggesting difference in the coupling to bosons mediating the Cooper pairs between relevant $d$ electron bands.
 
\end{abstract}


\maketitle


\section{Introduction}
The recent discovery of the iron pnictide superconductor
\lfaof\ (LFAO-F)  over a fluorine concentration of $0.05\le x\le 0.2$ 
with the maximal critical temperature ($T_{\rm c}$)
of 26~K \cite{R_Kamihara} and the following revelation of much increased
$T_{\rm c}$ upon the substitution of La for other rare-earth elements
(Ce, Pr, Nd, Sm,... leading to a maxium $T_{\rm c}$ of 55~K \cite{HCChen:08,ZARen:08,GFChen:08}) or the application of pressure for LFAO-F ($\sim$43~K \cite{R_Pressure}) have triggered broad interest in the mechanism yielding a relatively high $T_{\rm c}$ in this new class of compounds.   They have a layered structure like high-$T_{\rm c}$ cuprates, where the dopant and conducting layers are so separated that carriers ({\sl electrons}, in this case) introduced by the substitution of O$^{2-}$ with F$^{-}$ in the La$_2$O$_2$ layers move within the layers consisting of strongly bonded Fe and As atoms. Moreover, very recent developments demonstrate increasing variety in the methods of electron doping such as oxygen depletion \cite{Kito:08,ZARen:08-2} or Co substitution for Fe \cite{Sefat:08,Sefat:08-2,Qi:08,Matsuishi:08}.
A similar situation is presumed for the ternary compound $A$Fe$_2$As$_2$ ($A=$ Ba, Sr, Ca), where {\sl holes} are introduced by the substitution of $A^{2+}$ with $B^{+}$ ions ($B=$ Na, K, Cs) \cite{Rotter:08,Sasmal:08,GFChen:08-2,GWu:08}.  They exhibit another qualitative similarity to cuprates in that superconductivity occurs upon carrier doping of pristine compounds that exhibit magnetism \cite{Cruz:08,Nakai:08,Klauss:08,Carlo:08,McGuire:08,Kitao:08,Zhao:08,Aczel:08}.  Recent results of the muon spin rotation/relaxation ($\mu$SR) experiment on a variety of iron pnictide superconductors showed that the superfluid density $n_s$ may fall on the empirical line on the $n_s$ vs $T_{\rm c}$ diagram observed for the {\sl underdoped} cuprates \cite{Carlo:08,Luetkens:08}, from which possibility of the common mechanism of superconductivity is argued between oxypnictides and cuprates. 

The iron pnictides exhibit an interesting similarity with cuprates that the variation of $T_c$ against doping  is ``bell-shaped'' in hole-doped compounds \cite{Rotter:08,Sasmal:08} while $T_{\rm c}$ does not vary much with $x$ in electron-doped case \cite{R_Kamihara,Luetkens:08-2}.  However, recent investigations in electron-doped ($n$ type) cuprates strongly suggest that such an electron-hole ``asymmetry" is a manifestation of  difference in the fundamental properties of underlying electronic states between these two cases, where the $n$ type cuprates are much more like normal Fermi liquids rather than doped Mott insulators \cite{KSatoh:08}.  This might be readily illustrated by pointing out that, given all the doped carriers participate in the Cooper pairs (as suggested by experiment), the insensitivity of $T_{\rm c}$ against the variation of $n_s$ ($\propto x$) observed in LFAO-F \cite{R_Kamihara} cannot be reconciled with the above-mentioned empirical linear relation, while it is reasonably understood from the conventional BCS theory where condensation energy is predicted to be independent of carrier concentration. More interestingly, the very recent revelation of superconductivity upon Co substitution for Fe in LaFeAsO and other iron pnictides (where the Co atoms serves as electron donors) brings out a sheer contrast between these two classes of materials in terms of response to the substitution of transition metal ions as well as the tolerance of superconductivity to the distortion of conducting layers \cite{Sefat:08,Sefat:08-2,Qi:08,Matsuishi:08}.

The close relationship of magnetism and superconductivity suggests that a detailed investigation of how these two phases coexist (and compete) near the phase boundary will provide important clues to elucidating the paring mechanism. Among various techniques, $\mu$SR has a great advantage in that it can be applied in systems consisting of spatially inhomogeneous multiple phases, providing information on respective phases according to their fractional yield. Our $\mu$SR measurement in a LFAO-F sample with $x=0.06$ ($T_{\rm c}\simeq18$ K) reveals that these two phases indeed coexist in the form of macroscopic phase separation, and more interestingly, that a spin glass-like magnetic phase develops in conjunction with superconductivity in the paramagnetic phase \cite{Takeshita:08}.  This accordance strongly suggests a common origin of the electronic correlation between these two competing phases. On the other hand,  \msr\ study on \cfcaf\ (CFCAF, a variation of LFAO-F with trivalent cation and oxygen respectively replaced with divalent alkali metal and fluorine, and the carrier doping is attained by substituting Co for Fe) reveals a unique character of the Fe$_2$As$_2$ layers that the superconducting state is realized over a vicinity of Co atoms, as inferred from the observation that the superconducting volume fraction is nearly proportional to the Co concentration \cite{Takeshita:08-2}. The rest of the CFCAF specimen remains magnetic (strongly modulated spin density wave), thus indicating that superconductivity coexists with magnetism again in a form of phase separation. 

As already mentioned, \msr\ can provide information selectively from the superconducting parts of samples even in the situation that they coexist with magnetism. We examine the temperature dependence of $n_{\rm s}$ in LFAO-F, CFCAF, and in \bkfa\ (BKFA) \cite{Hiraishi:08}, and discuss the degree of anisotropy in their superconducting order parameters and the strength of coupling to bosons that mediate the Cooper pairs. 

\section{Experiment}

Although the oxypnictides $R$FeAsO$_{1-x}$F$_x$ with rare-earth substitution ($R=$ Nd, Sm, etc.) exhibits higher $T_{\rm c}$ than that of LFAO-F, strong random magnetic fields from rare-earth ions preclude a detailed study of the ground state using sensitive magnetic probes like \msr.
In this regard, the original LFAO-F (as well as other two compounds in this review) has a major advantage that we can attribute the origin of magnetism, if at all detected by \msr, to that of the Fe$_2$As$_2$ layers without ambiguity. The target concentration has been chosen near the phase boundary, $x=0.06$ for \lfaof\ and 0.075--0.15 for \cfcaf, where a polycrystalline sample has been synthesized by solid state reaction. The detailed procedure for sample preparation is described in an earlier report \cite{R_Kamihara,Matsuishi:08,Takeshita:08-2}.  For \bkfa, polycrystalline samples with nominal compositions of $x=0.1$ and 0.4 were prepared by a solid state reaction using starting materials having the highest purity available, where details of the preparation process are described in an earlier report\cite{GWu:08}.  The obtained LFAO-F sample was confirmed to be mostly of single phase using X-ray diffraction method (see figure \ref{dcChi}). Of two possible impurity phases, namely LaOF and FeAs, only the latter exhibits a magnetic (helical) order with $T_N\simeq77$ K \cite{Selte:72,Baker:08}. 
For the CFCAF samples, it was inferred from detailed X-ray diffraction analysis that
nearly 11\% of the sample was crystalized in Ca(Fe,Co)$_2$As$_2$ for the sample with $x=0.15$, while no such phase was observed for other samples. In addition, a small fraction (2--4\%) of CaF$_2$ (fluorite) was found to exist as impurity in all samples. Since muons in fluorite are depolarized immediately upon implantation, their contribution should be negligible. The influence of Ca(Fe,Co)$_2$As$_2$ will be discussed in the following section.   Finally, all the observed diffraction peaks in the \bkfa\ sample with $x=0.4$ was perfectly reproduced by those of single phase compound\cite{Rotter:08}, while a minor unknown impurity phase ($\sim$2\%) was observed for $x=0.1$. 

Conventional $\mu$SR measurement was performed using the
LAMPF spectrometer installed on the M20 beamline of TRIUMF, Canada. During the measurement under a zero field (ZF), residual magnetic field at the sample position was reduced below $10^{-6}$~T with the initial muon spin direction parallel to the muon beam direction
[$\vec{P}_\mu(0)\parallel \hat{z}$].  For longitudinal field (LF)
measurement, a magnetic field was applied parallel to $\vec{P}_\mu(0)$. Time-dependent muon polarization [$G_z(t)=\hat{z}\cdot \vec{P}_\mu(t)$] was monitored by measuring decay-positron asymmetry along the $\hat{z}$-axis.  Transverse field (TF) condition was realized by
rotating the initial muon polarization so that $\vec{P}_\mu(0)\parallel
\hat{x}$, where the asymmetry was monitored along the $\hat{x}$-axis to obtain $G_x(t)=\hat{x}\cdot \vec{P}_\mu(t)$.  All the measurements under a magnetic field were made by cooling the sample to the target
temperature after the field equilibrated.

\section{Result}
\subsection{\lfaof}
\begin{figure}[tp]
\begin{center}
\includegraphics[width=0.6\textwidth,clip]{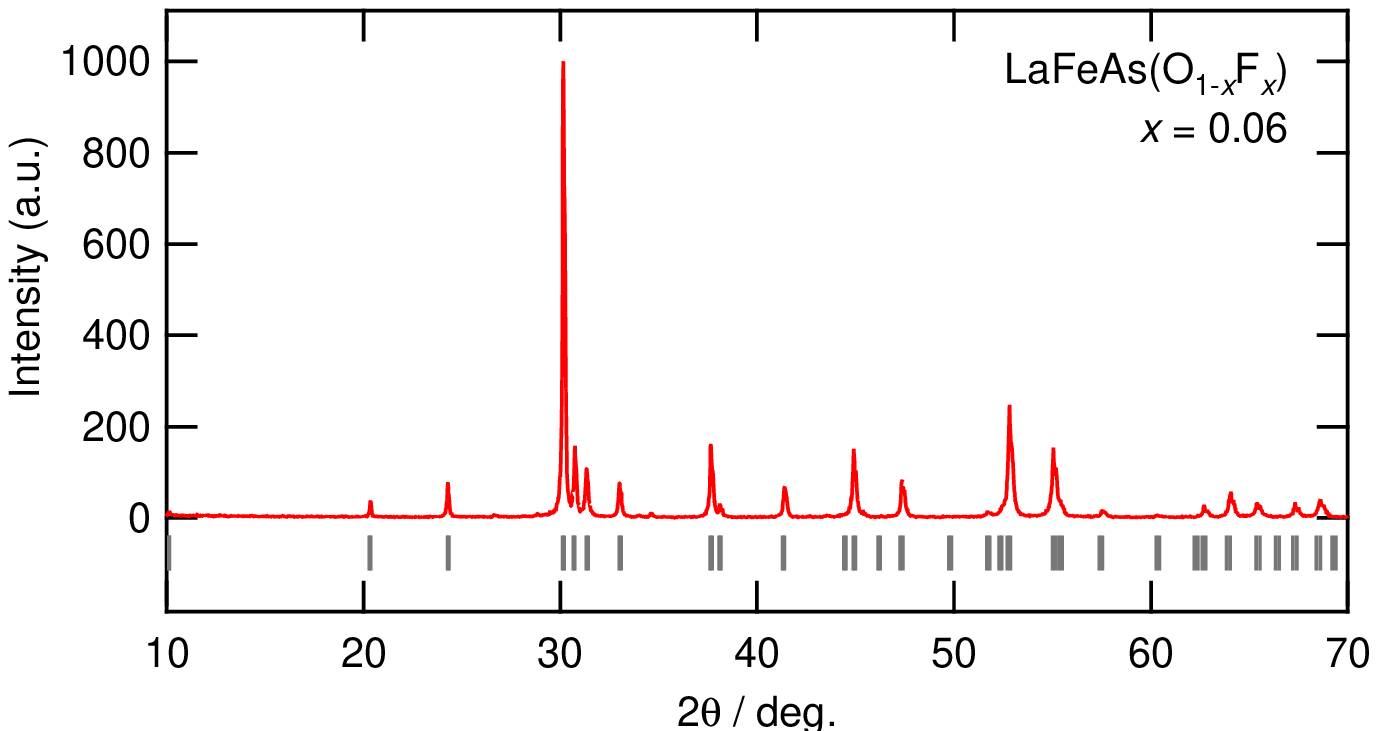}\\
\vspace{-3mm}
\includegraphics[width=0.6\textwidth,clip]{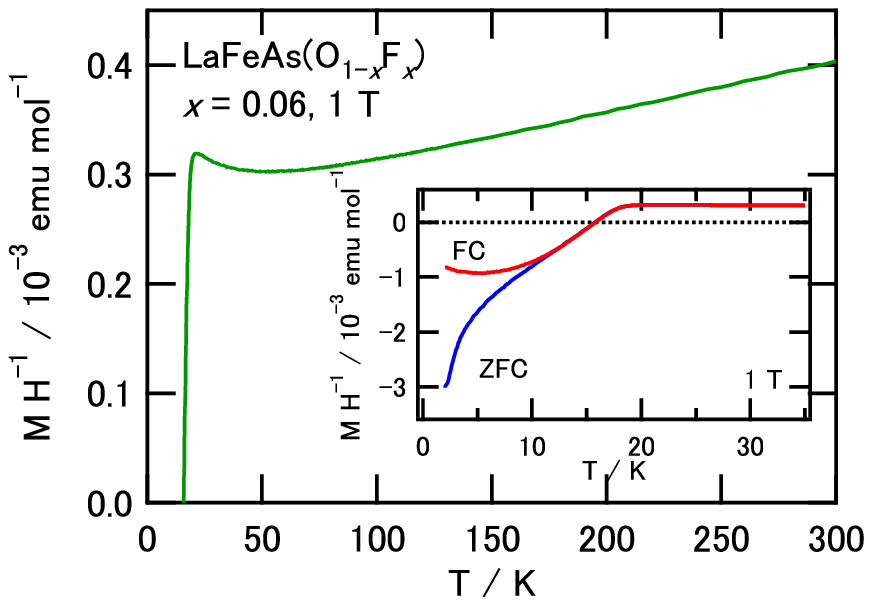}
\caption{(Color online)
X-ray diffraction (top) and Magnetic susceptibility data obtained 
for the sample of LaFeAsO$_{0.94}$F$_{0.06}$ measured by $\mu$SR measurement. Inset  shows a reduced view of the region below 35 K.}
\label{dcChi}
\end{center}
\end{figure}

\begin{figure}[tp]
\begin{center}
\includegraphics[width=0.5\textwidth,clip]{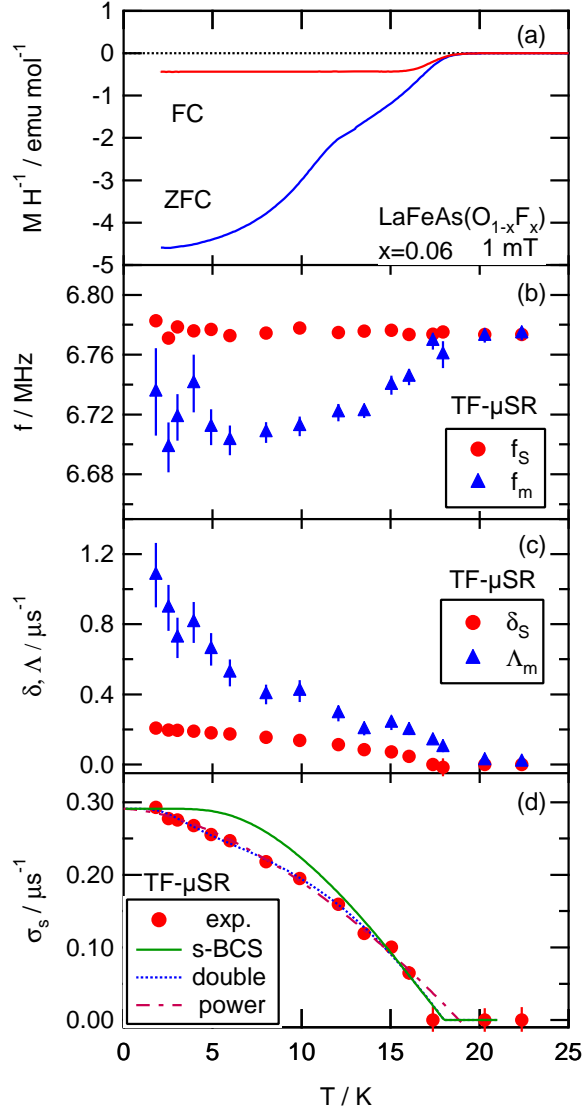}
\caption{(Color online)
Temperature dependence of dc magnetic susceptibility measured at 1 mT
 (a), and that of physical parameters deduced by analyzing TF-$\mu$SR
 spectra in superconducting ($f_{\rm s}$, $\delta_{\rm s}$) and magnetic
 ($f_{\rm m}$, $\Lambda_{\rm m}$) phases (b-c), and of $\sigma_{\rm s}$
 ($=\sqrt{2}\delta_{\rm s}$) proportional to superfluid density
 (d). Curves in (d) are fits by models described in the text.
}
\label{G_TFmulti}
\end{center}
\end{figure}

The quality of samples measured by \msr\ has been examined by looking into X-ray diffraction and magnetic susceptibility data.  As shown in figure \ref{dcChi}, magnetic susceptibility exhibits no trace of FeAs phase or local magnetic impurities except below $\sim50$ K where a small upturn is observed.  The susceptibility at a lower field [shown in figure~\ref{G_TFmulti}(a)] provides evidence of bulk superconductivity with $T_{\rm c}\sim18$~K from the onset of diamagnetism.  

\begin{figure}[tp]
\begin{center}
\includegraphics[width=0.6\textwidth,clip]{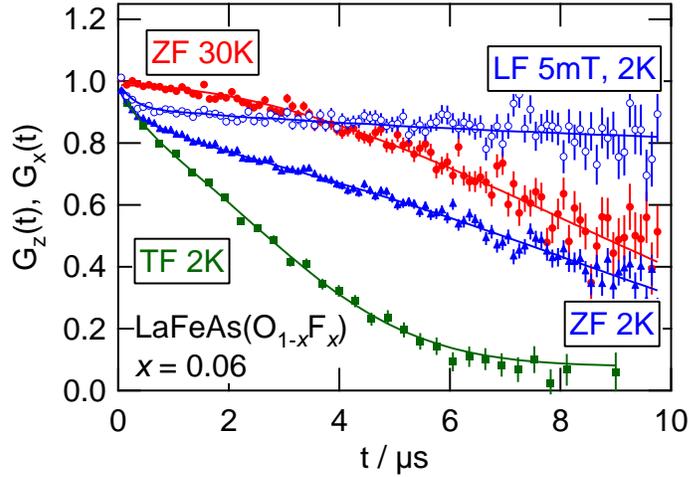}
\caption{(Color online)
$\mu$SR time spectra observed in \lfaof\ with
$x=0.06$ at 2 K under a longitudinal field (LF), a zero field (ZF), and
a transverse field (TF), and that under ZF at 30 K.
The spectrum under TF is plotted on a rotating reference
frame of 6.78 MHz to extract the envelop function. }
\label{G_TSAll}
\end{center}
\end{figure}

ZF-$\mu$SR is the most sensitive technique for examining magnetism in any form, where the development of local magnetic moments leads to either the spontaneous oscillation (for long-range order) or exponential damping (inhomogeneous local magnetism) of $G_z(t)$.  Figure \ref{G_TSAll} shows examples of ZF-$\mu$SR time spectra collected at 2~K and 30~K.  The spectrum at 30~K ($>T_{\rm c}$) exhibits a Gaussian-like depolarization due to weak random local fields from nuclear magnetic moments, indicating that the entire sample is in paramagnetic state.
Meanwhile, the spectrum at 2~K is split into two components, one that
exhibits a fast depolarization and the other that remains to show
Gaussian-like relaxation. This indicates that there is a finite fraction
of implanted muons that sense hyperfine fields from local electronic
moments.  The absence of oscillatory signal implies that the hyperfine
field is highly inhomogeneous, so that the 
local magnetism is characterized by strong randomness or spin fluctuation.  The fractional
yield of this component is as large as 25\% (see
below), which is hardly attributed to impurity and therefore implies
that the sample exhibits a macroscopic phase separation into two phases.

\par

The magnitude of the hyperfine field and that of spin fluctuation are
evaluated by observing the response of the $\mu$SR spectrum to a
longitudinal field (LF).  It is shown in figure~\ref{G_TSAll} that the slow 
depolarization of the paramagnetic component is quenched by applying a weak
magnetic field (LF=5 mT), which is perfectly in line with the
suppression of static nuclear dipolar fields ($<10^0$ mT).  Meanwhile,
the faster depolarization (seen for $0<t<1$ $\mu$s) due to the magnetic
phase is recovered only gradually over a field range of $10^{1\sim2}$
mT, and there still remains a slow depolarization even at the highest field
of 60 mT.  This residual depolarization under LF is a clear sign that
local spins are slowly fluctuating, leading to the spin-lattice
relaxation of $\vec{P}_\mu(t)$.  Such quasi-two-step depolarization is also
observed in dilute spin-glass systems,\cite{Uemura:81} which is
understood as a distribution of spin correlation time.  A detailed
analysis is made considering that these two components coming from the
magnetic phase (see below).

\par

Under a transverse field, implanted muons experience an inhomogeneity of
the field [$B_z({\bf r})$] due to flux line lattice formation below
$T_{\rm c}$ that leads to depolarization, in addition to those observed
under a zero field.  The TF-$\mu$SR time spectrum in figure~\ref{G_TSAll}
(envelop part of the oscillation) obtained under a field of 50 mT
exhibits complete depolarization at 2 K, indicating that the entire
volume of the paramagnetic phase falls into the superconducting state.
The rapidly depolarizing component observed under ZF is also visible
(although the coarse binning of the spectra for extracting the envelop
part makes it slightly obscure), indicating that the corresponding part
of the sample remains magnetic.

\begin{table}[tb]
\begin{center}
\begin{tabular}{cccc}
 \hline
$i$ &$w_i$ & $\delta_i$ ($\mu {\rm s}^{-1}$) 
& $\nu_i$ ($\mu {\rm s}^{-1}$)\\
 \hline
1  & 0.754(9)  &  -- & --\\
2 & 0.165(9) & 0.71(5) & 1.7(2)\\
3  &  0.081(4) & 3.9(3)  &  4(1)\\
 \hline
\end{tabular}
\caption{Physical parameters obtained from LF-$\mu$SR spectra 
at 2~K by analysis using eq.~(\ref{E_LF}).}
\label{T_LFPara}
\end{center}
\end{table}

\par

Considering the presence of the magnetic phase besides the paramagnetic
(=superconducting below $T_{\rm c}$) phase, we take special precaution
to analyze both TF and ZF/LF $\mu$SR spectra in a consistent manner.
For the determination of physical parameters describing the behavior of
signals from the magnetic phase, we first analyze ZF/LF spectra at 2~K
using the $\chi$-square minimization fit with the relaxation function
\begin{equation}
\label{E_LF}
G_z(t)=[w_1 + \sum^3_{i=2}w_i 
 \exp{(-\Lambda_i t)}]\cdot G_{\rm KT}(\delta_{\rm N}:t),
\end{equation}
where $G_{\rm KT}(\delta_{\rm N}:t)$ is the Kubo-Toyabe relaxation
function for describing the Gaussian damping due to random local fields
from nuclear moments (with $\delta_{\rm N}$ being the linewidth)
\cite{R_Hayano}, $w_1$ is the fractional yield for the paramagnetic
phase, $w_2$ and $w_3$ are those for the magnetic phase ($\sum w_i=1$) with
$\Lambda_{2, 3}$ being the corresponding relaxation rate described by
the Redfield relation
\begin{equation}
 \Lambda_{i} = \frac{2\delta_i^2 \nu_i}{\nu_i^2 + \omega_\mu^2}\label{lmdmg}
 \ \ \ (i=2,3),
\end{equation}
where $\omega_\mu=\gamma_\mu H_{\rm LF}$, $\gamma_\mu$ is the muon
gyromagnetic ratio ($=2\pi\times135.53$ MHz/T), $H_{\rm LF}$ is the
longitudinal field, $\delta_2$ and $\delta_3$ are the mean values of the
hyperfine fields being exerted on muons from local electronic moments, and
$\nu_2$ and $\nu_3$ are the fluctuation rates of the hyperfine fields.
The solid curves in figure~\ref{G_TSAll} show the result of analysis where
all the spectra at different fields (only ZF and LF=5mT are shown here)
are fitted simultaneously using eqs.~(\ref{E_LF}) and (\ref{lmdmg}) with
common parameter values (except $\omega_\mu$ that is fixed to the
respective value for each spectrum), which show excellent agreement with
all the spectra. The deduced parameters are shown in table~\ref{T_LFPara}. 
 Although the depolarization in the magnetic
phase is approximately represented by two components with different
hyperfine couplings ($\delta_i$), the fluctuation rates ($\nu_i$) are
close to each other (10$^7$ s$^{-1}$ at 2 K), suggesting that the
randomness is primarily due to the distribution in the size of local
moments (or in their distances to muons).  Since no impurity phase with
a fraction as large as 25\% is detected by X-ray diffraction analysis,
it is concluded that this magnetic phase is intrinsic.

\par
In the analysis of temperature-dependent TF spectra, we used the
relaxation function
\begin{equation}
\label{E_TFG}
 G_x(t)= [w_1e^{-\delta^2_{\rm s}t^2}\cos{(2\pi f_{\rm s} t+\phi)}\nonumber\\
  + (w_2+w_3)e^{-\Lambda_{\rm m} t}\cos{(2\pi f_{\rm m}t+\phi)}]e^{-\frac{1}{2}\delta^2_{\rm N}t^2},
\end{equation}
where $w_i$ and $\delta_{\rm N}$ are fixed to the values obtained by
analyzing ZF/LF-$\mu$SR spectra.  The first component in the above
equation represents the contribution of flux line lattice formation in
the superconducting phase, where $\delta_{\rm s}$ corresponds to the
linewidth $\sigma_{\rm s}=\sqrt{2}\delta_{\rm s}=\gamma_\mu\langle
(B({\bf r})-B_0)^2\rangle^{1/2}$ [with $B_0$ being the mean $B({\bf
r})$], while the second term represents the relaxation in the magnetic
phase. Here, the relaxation rate for the latter is represented by a
single value $\Lambda_{\rm m}$ (instead of $\Lambda_{2,3}$), as it turns
out that the two components observed under LF are hardly discernible in
TF-$\mu$SR spectra.  [This does not affect the result of the analysis,
because the amplitude is fixed to $w_2+w_3$ so that $\Lambda_{\rm m}$
may represent a mean $\simeq(w_2\Lambda_2+w_3\Lambda_3)/(w_2+w_3)$.]
The fit analysis using the above form indicates that all the spectra are
perfectly reproduced while the partial asymmetry is fixed to the value
determined from ZF-$\mu$SR spectra. This strengthens the presumption
that the paramagnetic phase becomes superconducting below $T_{\rm c}$.
The result of analysis is summarized in figure~\ref{G_TFmulti}, together
with the result of dc magnetization measured in the sample from the same
batch as that used for $\mu$SR.
\begin{table}[tp]
\begin{center}
\caption{Physical parameters deduced from curve fitting for the data in figure~\ref{G_TFmulti}(d).}
\begin{tabular}{c|cc||c|c}
\hline
 & Single-gap &Double-gap & \multicolumn{2}{|c}{Power law} \\
 \hline
  $T_{\rm c}$ (K)  &  18.0(4) &   18.0(5)   &$T_{\rm c}$ (K)  &  18.9(4)  \\
  $\sigma(0)$ ($\mu {\rm s}^{-1}$) & 0.273(2) & 0.291(5) 
 & $\sigma(0)$ ($\mu {\rm s}^{-1})$ & 0.291(4)\\
  $2\Delta_1/k_{\rm B}T_{\rm c}$   &  $3.3(1)$   &  4.2(4) & $\beta$    &  1.7(1) \\
  $2\Delta_2/k_{\rm B}T_{\rm c}$   &   --   &  1.1(3) &  & \\
 $w$  &  --                &  0.73(6)  & & \\
 \hline
$\chi^2/N_f$ & 5.19 &  0.87 & $\chi^2/N_f$ & 1.22 \\
 \hline
\end{tabular}
\label{T_SgmPara}
\end{center}
\end{table}

 \par

It is interesting to note in figure~\ref{G_TFmulti}(b) that, although the
central frequency in the superconducting phase ($f_{\rm s}$) does not
show much change below $T_{\rm c}\simeq18$ K probably owing to a large
magnetic penetration depth (it is indeed large, see below), that in the
magnetic phase ($f_{\rm m}$) exhibits a clear shift in the negative
direction below $T_{\rm m}\simeq T_{\rm c}$. The magnitude of the shift
is as large as $\sim1$\% and thus is readily identified despite a
relatively low external field of 50 mT.  As shown in
figure~\ref{G_TFmulti}(c), the relaxation rate in the magnetic phase
($\Lambda_{\rm m}$) also develops below $T_{\rm c}$ in accordance with
the frequency shift, demonstrating that a spin-glass-like magnetism sets
in below $T_{\rm c}$.  Here, we note that the development of magnetic
phase is already evident in the ZF/LF-$\mu$SR spectra, and results are
fully consistent with each other.  The onset of superconductivity below
$T_{\rm c}$ is also confirmed by an increase of $\delta_{\rm s}$, as
observed in figure~\ref{G_TFmulti}(c).  This remarkable accordance of
onset temperature between magnetism and superconductivity strongly
suggests that there is an intrinsic relationship between the
superconducting and magnetic phases that leads to a common
characteristic temperature.

\par

The temperature dependence of $\sigma_{\rm s}$ in
figure~\ref{G_TFmulti}(d) is compared with theoretical predictions for a
variety of models with different order parameters. The weak-coupling BCS
model ($s$-wave, single gap) apparently fails to reproduce the present
data, as they exhibit a tendency to vary with temperature over the
region $T/T_{\rm c}<0.4$.  
As a next possibility, we compare the data with a phenomenological double-gap
model\cite{R_TwoGap,Ohishi:03},
\begin{eqnarray}
\sigma_{\rm s}(T)&=&\sigma(0)-w{\cdot}\delta\sigma(\Delta_1,T)-(1-w){\cdot}\delta\sigma(\Delta_2,T),  \label{dblgap} \\
\delta\sigma(\Delta,T)&=& \frac{2\sigma(0)}{k_{\rm B}T}\int_0^{\infty}f(\varepsilon,T){\cdot}[1-f(\varepsilon,T)]d\varepsilon, \nonumber \\
f(\varepsilon,T) &=& \left(1+e^{\sqrt{\varepsilon^2+\Delta(T)^2}/k_{\rm B}T}\right)^{-1},\nonumber 
\end{eqnarray}
where $\Delta_i$ ($i$ = 1 and 2) are the energy gap at $T=0$, $w$ is the relative weight for $i=1$, $k_{\rm B}$ is the Boltzmann constant, $f(\varepsilon,T)$ is the Fermi distribution function, and $\Delta(T)$ is the standard BCS gap energy. 
A fit using the above model shown by a dotted curve seems to exhibit reasonable
agreement with data (the value of $2\Delta_1/k_{\rm B}T_{\rm c}$ in ref.\cite{Takeshita:08} is incorrect and it should be replaced with that in table 2).  These observations suggest that the superconducting order parameter is not described by
a s-wave symmetry with single gap.  When a power law, $\sigma_{\rm s} =
\sigma_0 [1-(T/T_{\rm c})^\beta]$, is used in fitting the data, we
obtain a curve shown by the broken line in figure~\ref{G_TFmulti}(d) with
an exponent $\beta\simeq2$.  This seems to favor the recent theory of fully gapped anisotropic 
$\pm s$-wave superconductivity that predicts more enhanced quasiparticle excitations
associated with the smaller gap \cite{Nagai:08}.

\par

In the limit of extreme type II superconductors [i.e.,
$\lambda/\xi\gg1$, where $\lambda$ is the effective London penetration
depth and $\xi=\sqrt{\Phi_0/(2\pi H_{\rm c2})}$ is the Ginzburg-Landau
coherence length, $\Phi_0$ is the flux quantum, and $H_{\rm c2}$ is the
upper critical field], $\sigma_{\rm s}$ is determined by $\lambda$ using
the relation \cite{R_Brandt} 
\begin{equation}
 \sigma_{\rm s}/\gamma_\mu=
2.74\times10^{-2}(1-h)\left[1+3.9(1-h)^2\right]^{1/2}\Phi_0\lambda^{-2},\label{sgmh}
\end{equation}
where $h=H_{\rm TF}/H_{\rm c2}$ and $H_{\rm TF}$ is the magnitude of
external field. While eq.~(\ref{sgmh}) is a good approximation under limited conditions (as it introduces a relatively large cutoff $\sim\sqrt{2}/\xi$ in calculating the second moment, see ref.\cite{Brandt:03} for more detail), we would continue using it to keep the consistency with earlier analysis. 
 From $\sigma_{\rm s}$ extrapolated to $T=0$ and taking
$H_{\rm c2}\simeq50$~T (ref.~\cite{R_Hc2}), we obtain
$\lambda$=595(3)~nm.  Because of the large anisotropy expected from the
layered structure of this compound, $\lambda$ in the polycrystalline
sample would be predominantly determined by in-plane penetration depth
($\lambda_{\rm ab}$).  Using the formula $\lambda=1.31\lambda_{\rm ab}$
for such a situation \cite{R_Fesenko}, we obtain $\lambda_{\rm
ab}$=454(2)~nm.  This value coincides with that expected from the
aforementioned empirical relation between $\lambda^{-2}_{\rm ab}$
superconductors \cite{R_Uemura,Carlo:08}.  However, this may not be
uniquely attributed to the superfluid density because $\lambda$ depends
not only $n_{\rm s}$ but also on the effective mass, $ \sigma_{\rm
s}\propto\lambda^{-2}= n_{\rm s}e^2/m^*c^2$.

\par

Finally, we discuss the feature of the spin glass-like phase.  Assuming
that the local moments are those of off-stoichiometric iron atoms with a
moment size close to that in the SDW phase ($\sim0.36\mu_B$
\cite{Cruz:08}), the mean distance between muon and iron moments in the
relevant phase is estimated to be $\sim0.5$ nm from an average of
$\delta_i$.  Given the unit cell size ($a=0.403$ nm, $c=0.874$ nm
\cite{R_Kamihara}), this would mean that more than a quarter of iron
atoms in the magnetic phase (i.e., $\simeq7$\% of the entire sample)
should serve as a source of local moments. It is unlikely that such a
significant fraction of iron atoms remains as impurities in the present
sample.

It might also be noteworthy that there is an anomaly near $T_{\rm
m2}\simeq12$ K in the susceptibility [the onset of ZFC/FC hysteresis in
figure~\ref{dcChi} and a steplike kink in figure~\ref{G_TFmulti}(a)].  This
seems to be in accordance with the onset of a steeper increase in
$\Lambda_{\rm m}$ below $T_{\rm m2}$, suggesting a change in magnetic
correlation.

\subsection{\cfcaf}

\begin{figure}[tp]
\begin{center}
\includegraphics[width=0.5\textwidth,clip]{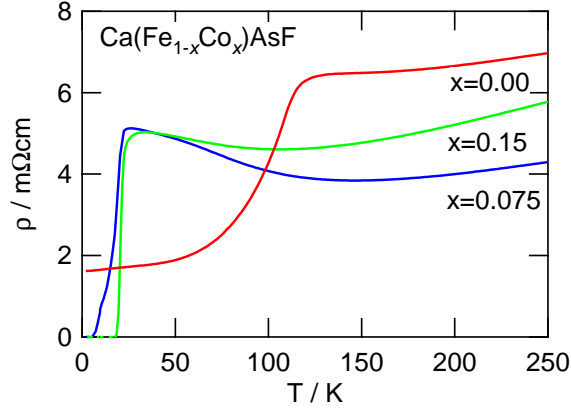}
\caption{(Color online)
Temperature dependence of electrical resistivity in \cfcaf\ with $x=0$, 0.075, and 0.15.
}
\label{cf-rho}
\end{center}
\end{figure}
\begin{figure}[tp]
\begin{center}
\includegraphics[width=0.6\textwidth,clip]{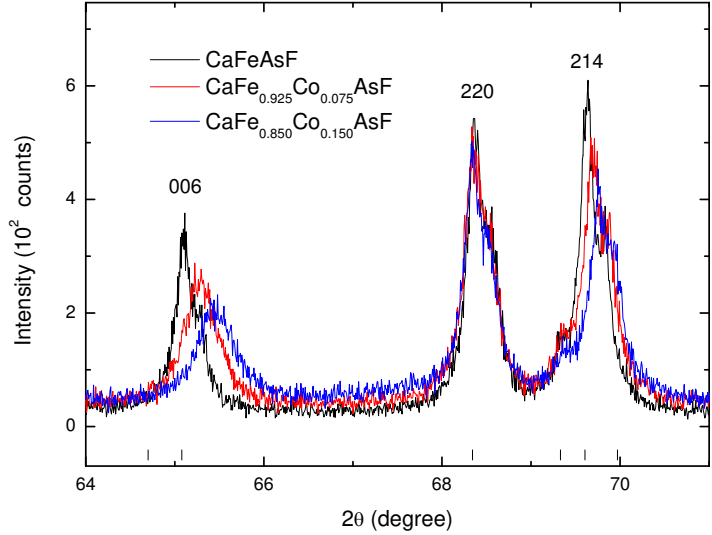}
\caption{(Color online)
X-ray diffraction pattern observed in \cfcaf\ with $x=0$, 0.075, and 0.15.
The [006] and [214] peaks exhibit a progressive shift with doping concentration, indicating that the majority of the samples consist of a uniform solid-solution.}
\label{cf-xray}
\end{center}
\end{figure}
\begin{table}[tp]
\begin{center}
\caption{Result of X-ray diffraction analysis on \cfcaf\ samples.}

\begin{tabular}{cc|cc|cc}
\hline
 \multicolumn{2}{c|}{$x=0$} &  \multicolumn{2}{c|}{$x=0.075$} &  \multicolumn{2}{c}{$x=0.15$}  \\
 \hline
CaFeAsF & 92.50\% & Ca(Fe,Co)AsF & 98.60\% &  Ca(Fe,Co)AsF & 85.74\% \\
CaF$_2$ & 2.24\% & CaF$_2$  & 1.40\% & CaF$_2$  & 3.71\% \\
FeAs & 5.26\% & --- & ---  &  Ca(Fe,Co)$_2$As$_2$ & 10.54\% \\
 \hline
\end{tabular}
\label{cf-xray-ana}
\end{center}
\end{table}

\msr\ measurements has been made on samples with $x=0$, 0.075, and 0.15, where the pristine compound exhibits anomaly around 120 K in the resistivity while the latter two fall into superconducting state below $T_c\simeq18$ K and 21 K (defined as the midpoint of fall in the resistivity, see figure~\ref{cf-rho}) \cite{Matsuishi:08}.  The sample quality is examined by detailed X-ray diffraction analysis. As shown in figure~\ref{cf-xray}, the peaks representing the Fe$_2$As$_2$ layers exhibit a progressive shift with Co content, which is consistent with the formation of uniform solid-solution.  Meanwhile, a slight broadening 
of the [006] peak may suggest the fluctuation of lattice constants due to randomness introduced by Co substitution.
The impurity phases deduced from X-ray diffraction analysis is summarized in table \ref{cf-xray-ana}, where it is noticeable that nearly 11\% of the sample with $x=0.15$ is crystalized in Ca(Fe,Co)$_2$As$_2$ (Ca122 phase), while no such phase was observed for other samples. It is known that CaFe$_2$As$_2$ is antiferromagnetic below $T_N=171$ K, where the transition is associated with structural change to orthorhombic \cite{Park:08}.  So far there seems to be no report on the electronic property upon Co substitution for this compound. We therefore presume that muons stopped in the Ca122 phase does not contribute to the \msr\ signal below 171 K.

\begin{figure}[tp]
\begin{center}
\includegraphics[width=0.6\textwidth,clip]{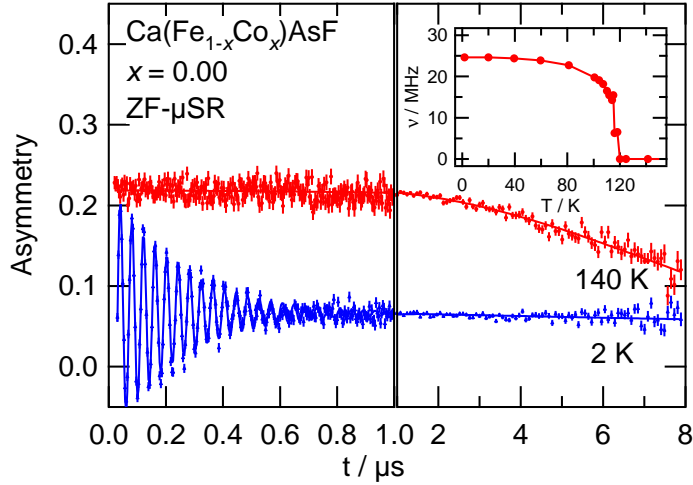}
\caption{(Color online)
ZF-\msr\ time spectra observed in CaFeAsF (undoped) at 140 K and 2 K. 
(inset) Frequency of spontaneous oscillation ($\nu\equiv f_{\rm m}$) vs temperature.
}
\label{cf-zf0}
\end{center}
\end{figure}

 It is inferred from ZF-\msr\ measurement that the anomaly observed in the sample with $x=0$ corresponds to the occurrence of magnetic phase below $T_{\rm m}\simeq120$ K. As shown in figure~\ref{cf-zf0}, the \msr\ spectra below $T_{\rm m}$ exhibit a spontaneous oscillation with a well defind frequency that approaches $f_{\rm m}\simeq25$ MHz with decreasing temperature.  This is an indication that implanted muons sense a unique internal magnetic field $B_{\rm m}=2\pi f_{\rm m}/\gamma_\mu\simeq0.18$ T.  The magnitude of $B_{\rm m}$ is in good agreement with earlier \msr\ results in $R$FeAsO \cite{Klauss:08,Carlo:08,Aczel:08}, where a commensurate spin density wave (SDW) with a reduced moment of $\sim$0.25$\mu_B$ at the iron sites is suggested \cite{Klauss:08}. It is also inferred from LF-\msr\ spectra that the internal field is static within the time scale of \msr\ ($<10^{-5}$ s). 

\begin{figure}[tp]
\begin{center}
\includegraphics[width=0.95\textwidth,clip]{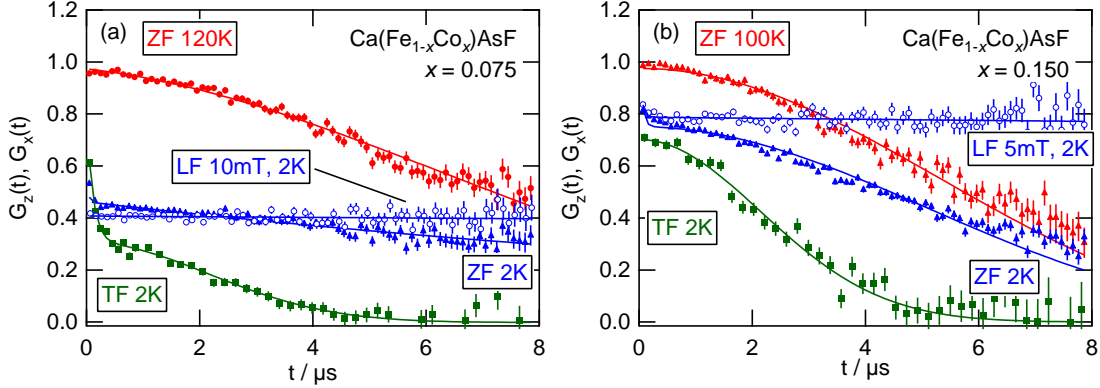}
\caption{(Color online) $\mu$SR time spectra observed in \cfcaf\ [$x=0.075$ (a) and 0.15 (b)] at 2 K under a longitudinal field (LF, open circles), a zero field (ZF, triangles), and a transverse field (TF, squares), and that under ZF above $T_m$ (triangles). The spectrum under TF is plotted on a rotating reference
frame to extract the envelop function.
}
\label{cf-tsp}
\end{center}
\end{figure}

It has been reported that the Co doping is quite effective to suppress the anomaly in the resistivity at $T_{\rm m}$; it virtually disappears at $x\simeq 0.1$ where the superconductivity seems to be close to its optimum as suggested from the maximal $T_c\simeq22$ K \cite{Matsuishi:08}.  However, ZF-\msr\ measurements in the samples with $x=0.075$ and 0.15 indicates that the superconductivity does not develop uniformly over the specimen. As shown in figure~\ref{cf-tsp}, the time spectra exhibit a character similar to those observed in LFAO-F ($x=0.06$), namely, they consist of two components, one showing rapid depolarization and other showing slow Gaussian damping with the relative yield of the latter increasing progressively with $x$.  A closer look into the earlier time range of the spectra obtained for $x=0.075$ indicates that the rapid depolarization corresponds to a strongly damped oscillation with a frequency roughly equal to $f_{\rm m}$.  This, together with the common onset temperature for magnetism ($T_{\rm m}\simeq120$ K), confirms that the signal comes from the SDW phase with a strong modulation due to Co doping.  Very recent neutron diffraction experiment suggests a similar situation for $x=0.06$, although they only observe a volume-averaged signal \cite{Xiao:08}. As inferred from TF-\msr\ measurements (see below), the rest of the specimen exhibits superconductivity below $T_c$.  

\begin{figure}[tp]
\begin{center}
\includegraphics[width=0.6\textwidth,clip]{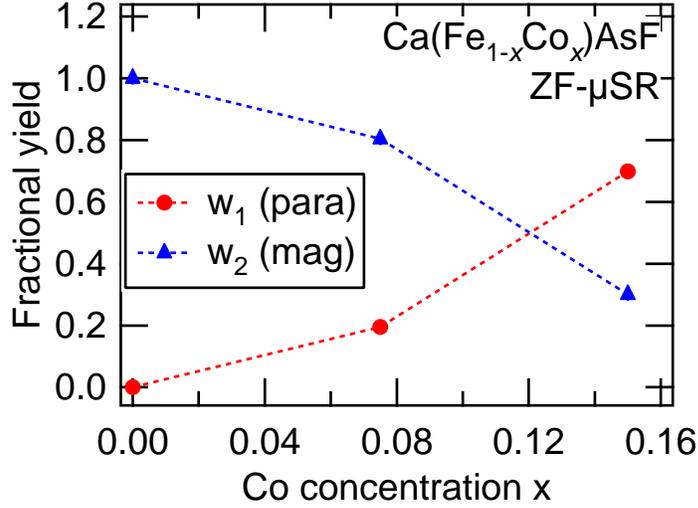}
\caption{(Color online) Relative yield of volumetric fraction for superconducting ($w_1$) and magnetic ($w_2$)   components in \cfcaf\ as a function of Co concentration ($x$).
}
\label{cf-frac}
\end{center}
\end{figure}

Considering these features, ZF-\msr\ spectra are analyzed by the $\chi$-square minimization fit with the relaxation function 
\begin{eqnarray}
G_z(t) &=&\left[w_1+w_2G_{\rm m}(t)\right]G_{\rm KT}(\delta_{\rm N}:t),\\ \label{E_ZF2}
& & G_{\rm m}(t)= \frac{1}{3} + \frac{2}{3}e^{-\Lambda_{\rm m} t}\cos(2\pi f_{\rm m} t+\phi),\label{E_mag}
\end{eqnarray}
where  $w_1$ is the fractional yield for the nonmagnetic phase, and $w_2$ is that for the SDW phase ($\sum w_i=1$) with $\Lambda_{\rm m}$ being the relaxation rate for the spontaneous oscillation. The first term in the second component represents the spatial average of $\cos\theta$ with $\theta$ referring to the angle between direction of the initial muon polarization and that of the {\sl static} local field at the muon site, which equals 1/3 in polycrystalline specimen under zero external field. (This term would be also subject to depolarization in the case of fluctuating local field, as observed in LFAO-F). The fractional yields of respective components are shown in figure~\ref{cf-frac}. 

\begin{figure}[tp]
\begin{center}
\includegraphics[width=0.95\textwidth,clip]{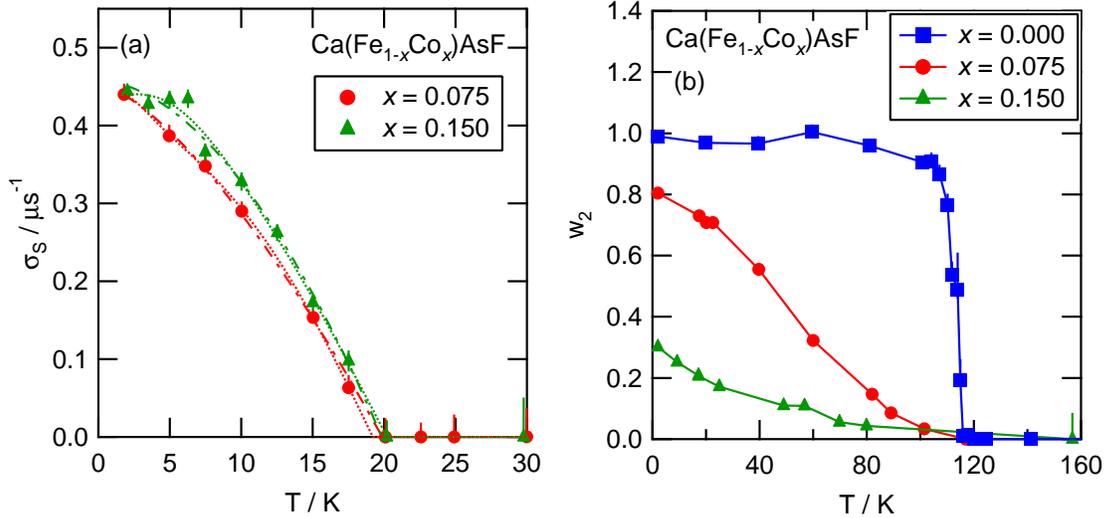}
\caption{(Color online)
Temperature dependence of (a) superfluid density ($\sigma_{\rm s}=\sqrt{2}\delta_{\rm s}$) and  (b) fractional yield of magnetic phases ($w_2$) for $x=0,$ 0.075  and 0.15.  Curves in (a) are fits by models described in the text.
}
\label{G_TFmulti2}
\end{center}
\end{figure}

The superconducting properties are extracted from TF-\msr\ spectra by fitting analysis in a manner similar to the case of LFAO-F, where we use a simplified form of eq.~(\ref{E_TFG})
with the factor $w_2+w_3$ replaced with $w_2$,
\begin{equation}
 G_x(t)= [w_1e^{-\delta^2_{\rm s}t^2}\cos(2\pi f_{\rm s} t+\phi)
 +w_2G_{\rm m}(t)]e^{-\frac{1}{2}\delta^2_{\rm N}t^2}.\label{E_TFGc}
\end{equation}
  It is confirmed that the spectra are
perfectly reproduced while the partial asymmetry ($w_i$) is fixed to the value determined from ZF-$\mu$SR spectra, supporting the presumption that the paramagnetic phase becomes superconducting below $T_{\rm c}$. The result of analysis is summarized in figure~\ref{G_TFmulti2}.

Unlike the case of LFAO-F, the magnetic component develops at temperatures much higher than the superconducting transition temperature ($T_{\rm m}>T_{\rm c}$), although $T_{\rm m}$ is considerably reduced to $\sim$80 K for $x=0.15$ (see figure~\ref{G_TFmulti2}).  
Meanwhile, the oscillation observed in figure~\ref{cf-zf0} disappears in both of the Co-doped samples (figure~\ref{cf-tsp}, ZF, 2K), indicating that the magnetic order is strongly modulated. This observation provides evidence that the phase separation is not simply due to aggregation of cobalt atoms upon sample preparation.  

As shown in figure~\ref{G_TFmulti2}, $\sigma_{\rm s}$ ($\propto n_{\rm s}$) is almost independent of $x$, including its temperature dependence.  Considering that the volume fraction for superconducting phase ($w_1$) is nearly proportional to $x$, this insensitivity of $n_{\rm s}$ to $x$ indicates that the superfluid (and corresponding carrier density in the normal state) stays in certain domains (``islands") centered at Co ions. A crude estimation suggests that the domain size may be $d_s\sim(abc/2\cdot0.8/0.15)^{1/3}\simeq0.9$ nm in diameter (where $a$, $b$, and $c$ are  unit cell size).  In other words, the superfluid behaves as an incompressible fluid in CFCAF.  

\begin{table}[tp]
\begin{center}
\caption{Physical parameters deduced from curve fitting for the data in figure~\ref{G_TFmulti2}(a).}
\begin{tabular}{c|cc|cc||c|cc}
\hline
& \multicolumn{2}{|c|}{Single-gap} & \multicolumn{2}{|c||}{Double-gap}  & \multicolumn{3}{c}{Power law} \\
 \hline
 $x$ & 0.075 & 0.15 & 0.075 & 0.15 & $x$ & 0.075 & 0.15 \\
 \hline
 $T_{\rm c}$ (K)  &  20.5(8) & 20.6(6)  & 19.2(9) &20.2(5) & $T_{\rm c}$ (K)    & 19.9(2)  & 19.8(3)\\
 $\sigma(0)$ ($\mu {\rm s}^{-1}$) & 0.42(1) &  0.44(1) & 0.44(1) & 0.44(1)  & $\sigma(0)$ ($\mu {\rm s}^{-1})$ & 0.45(1) & 0.46(1) \\
 $2\Delta_1/k_{\rm B}T_{\rm c}$  & 2.4(1)  & 2.9(1)  &  3.7(8)  & 3.9$\pm$3.1& $\beta$    &  1.5(1) &  1.8(1)\\
 $2\Delta_2/k_{\rm B}T_{\rm c}$  & -- &  -- &  1.1(7) & 2.3$\pm$1.5 & & & \\
 $w$    & --  &   --  & 0.73(18)   & 0.5$\pm$1.3  & \\
 \hline
 $\chi^2/N_f$  & 0.92 &  1.10 &  0.063 &  1.39  & $\chi^2/N_f$ & 0.27  & 1.91   \\
 \hline
\end{tabular}
\label{T_SgmPara2}
\end{center}
\end{table}

Following LFAO-F, the temperature dependence of $\sigma_{\rm s}$ in
figure~\ref{G_TFmulti2} is compared with theoretical predictions for a
variety of models with different order parameters. The weak-coupling BCS
model ($s$-wave, single gap) again fails to reproduce the present
data for both cases of $x=0.075$ and 0.15, as they exhibit a tendency to vary with temperature over the region $T/T_{\rm c}<0.4$.  A fit using two-gap model [eq.~(\ref{dblgap})] shown by solid curves exhibit reasonable
agreement with data with parameters listed in table~\ref{T_SgmPara2}, although the deduced gap parameters are not necessarily consistent with the prediction of the weak-coupling BCS model (as the larger gap has a ratio $2\Delta_1/k_{\rm B}T_{\rm c}$ greater than the BCS prediction of 3.53). 
In such a situation, one might be concerned about the effect of flux pinning that often leads to a distortion of temperature dependence of $\sigma_{\rm s}$.  However, we found that this would not be the case for the present samples, as inferred from the fidelity of $f_s$ upon sweeping external field.
These observations suggest that the superconducting order parameter is not described by a s-wave symmetry with single gap.   When a power law, $\sigma_{\rm s} =
\sigma_0 [1-(T/T_{\rm c})^\beta]$, is used in fitting the data, we
obtain result shown by the broken curves in figure~\ref{G_TFmulti2} with respective exponents $\beta\simeq1.5(1)$ for $x=0.075$ and 1.8(1) for  $x=0.15$.  As shown in the previous section,  a very similar behavior is observed in LFAO-F near the phase boundary ($x=0.06$) \cite{Takeshita:08}, which is consistent with the recent theory of fully gapped anisotropic 
$\pm s$-wave superconductivity \cite{Nagai:08}.

\subsection{\bkfa}
\begin{figure}[bp]
\begin{center}
\includegraphics[width=0.5\textwidth,clip]{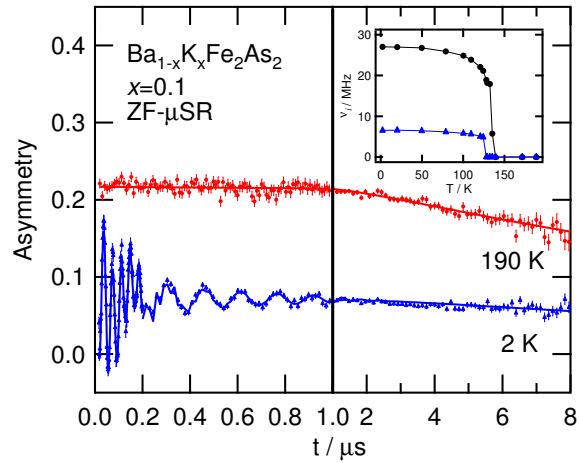}
\caption{(Color online)
ZF-$\mu$SR time spectra observed at 190 K and 2 K in \bkfa\ sample with $x=0.1$ ($T_c<2$K), where a spontaneous muon precession signal (mainly consisting of two frequencies) is clearly seen. Inset shows temperature dependence of the precession frequency. }
\label{bfa-tsp}
\end{center}
\end{figure}

In contrast to the case of preceding two families of compounds, the substitution of divalent barium with monovalent potassium in \bkfa\ leads to {\sl hole} doping over the Fe$_2$As$_2$ layers. Microscopic evidence shows that the parent compound BaFe$_2$As$_2$ exhibits magnetic order (SDW) below 140 K which is accompanied by a structural phase transition \cite{Rotter:08-2,Aczel:08}. The situation is common to Sr$_{1-x}$K$_x$Fe$_2$As$_2$ \cite{Zhao:08}, and it suggests that the electronic ground state of Fe$_2$As$_2$ layers in the parent compound is quite similar to that in LaFeAsO.  However,  the doping phase diagram is markedly different from LFAO-F, as it exhibits superconductivity over a wide range of hole content $p=x/2$ (per FeAs chemical formula) from 0.05 to 0.5 that far exceeds LFAO-F (i.e., $0.06\le x\le 0.2$)\cite{Sasmal:08}.  The phase diagram is also characterized by a bell-shaped variation of $T_c$ against $x$, where the maximal $T_c\simeq38$ K is attained near $x\sim0.4$ ($p\sim0.2$).  Considering the manifold nature of electronic band structure of Fe$_2$As$_2$ layers suggested by theories\cite{Mazin:08,Kuroki:08}, this difference may be attributed to that of the bands relevant to the doped carriers.  Our \msr\ result suggests that, while the magnetic ground state for $x=0.1$ is very similar to that of other pristine iron pnictides, the superconducting property in the optimally-doped sample is characterized by gap parameters that are much greater than  those of LFAO-F and CFCAF \cite{Hiraishi:08}. 

Figure \ref{bfa-tsp} shows the \msr\ spectra obtained upon muon implantation to the sample with $x=0.1$ under zero external field, where one can readily identify an oscillatory signal with multiple frequency components in the spectrum at 2 K.  The Fourier transform of the spectrum indicates that there are actually two of them, one approaching to 27 MHz and another to 7 MHz (corresponding to  internal fields of 0.2 and 0.05 T, respectively).  This is qualitatively similar to the situation in LaFeAsO \cite{Klauss:08,Carlo:08}, and thereby suggests that the high frequency component corresponds to the signal from muons situated on the Fe$_2$As$_2$ layers while another coming from those located near the cation sites. The onset temperature for the high frequency component is close to 140 K, which is also consistent with earlier reports \cite{Rotter:08-2,Aczel:08}.

\begin{figure}[tp]
\begin{center}
\includegraphics[width=0.5\textwidth,clip]{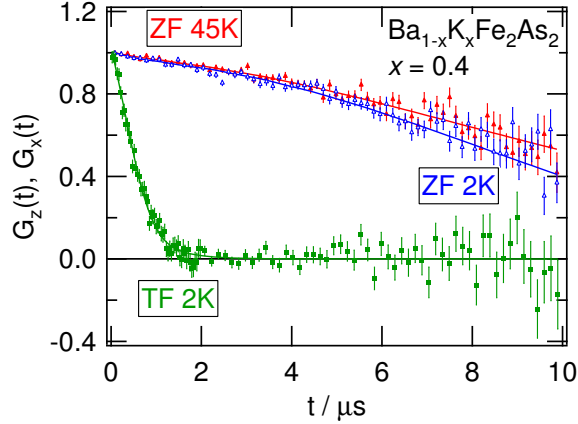}
\caption{(Color online)
 ZF-$\mu$SR time spectra observed at 45 K and 2 K in \bkfa\ sample with $x=0.4$ ($T_c\simeq38$ K). No trace of magnetic phase is observed. The spectra under a transverse field (TF) at 2 K is displayed on a rotating reference frame to extract the envelop function. Solid curves are fits using a Gaussian relaxation function described in the text. }
\label{bkfa-tsp}
\end{center}
\end{figure}

\begin{figure}[tp]
\begin{center}
\includegraphics[width=0.5\textwidth,clip]{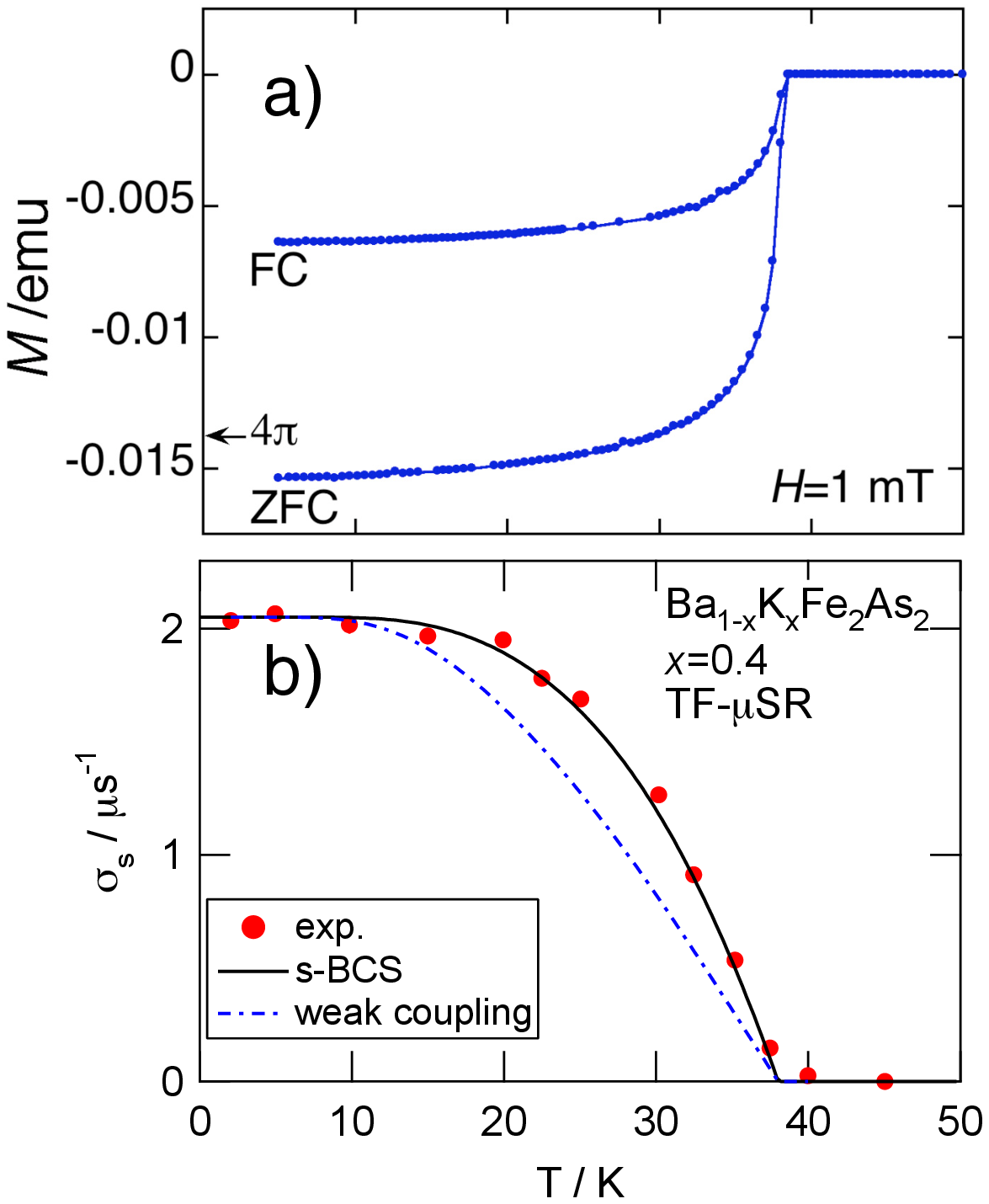}
\caption{(Color online)
a) Magnetization of \bkfa\ measured on the \msr\ sample with $x=0.4$, where data were obtained after cooling under an external field (FC) or zero field (ZFC). The total weight of the sample is 0.1 g, and the magnetization corresponding to $4\pi$ is obtained by using the structural parameters reported in Ref.\cite{Rotter:08}.
b) Temperature dependence of Gaussian linwidth $\sigma_{\rm s}=\sqrt{2}\delta_{\rm s}$  determined by TF-\msr\ measurement with $H=0.1$ T.  Curves represent the prediction of weak coupling BCS theory, where solid curve is a result of fit with a free gap parameter ($2\Delta/k_BT_c$), and
broken curve is that with $2\Delta/k_BT_c$ fixed to 3.53 (predicted by the BCS theory).}
\label{bkfa-chi}
\end{center}
\end{figure}

On the other hand, it is inferred from ZF-\msr\ spectra in figure~\ref{bkfa-tsp} that no trace of magnetism is found in the sample with $x=0.4$. This is in marked contrast with the earlier report on \msr\ measurements in a sample with $x=0.45$ where a magnetic phase seems to dominate over a large volume fraction \cite{Aczel:08}.  The quality of our specimen can be assessed by looking into bulk magnetization data, which is shown in figure~\ref{bkfa-chi}(a).  The sharp onset as well as a large Meissner fraction ($>4\pi$ for ZFC) indicates excellent quality of the specimen.  Assuming a general value for the Fermi velocity ($\sim10^6$ m/s), the mean free path ($l$) of carriers estimated from resistivity just above $T_c$ is about 3.5 nm. Since this figure should be regarded as a lower boundary (as it is determined near $T_c$), the actual $l$ must be much greater than the coherence length at lower temperatures ($\xi\simeq2.1$ nm, deduced from $H_{c2}\ge75$ T\cite{Ni:08}). Thus, we can conclude that the specimen is in the clean limit, and that the anisotropy in the superconducting order parameter, if at all, should be reflected in the temperature dependence of superfluid density measured by \msr.

The \msr\ time spectra under a transverse field of 0.1 T is shown in figure~\ref{bkfa-tsp}, where the envelop of the damping oscillation is extracted.  It exhibits depolarization towards zero, indicating that the entire sample falls into the flux line lattice state to exert strongly inhomogeneous internal field to implanted muons. The lineshape is well represented by a Gaussian damping, and the analysis is made by curve fit using a further simplified version of eq.~(\ref{E_TFGc}) with $w_2$ set to zero, namely,
\begin{equation}
\label{E_TFG2}
 G_x(t)=\exp\left[-\frac{1}{2}(\delta^2_{\rm N}+\sigma^2_{\rm s})t^2\right]\cos{(2\pi f_{\rm s} t+\phi)},
\end{equation}
where $\delta^2_{\rm N}$ is determined by fitting data above $T_c$ and subtracted from the total linewidth for the spectra below $T_c$.

The deduced linewidth, $\sigma_{\rm s}$, is plotted against temperature in figure~\ref{bkfa-chi}(b), where $\sigma_{\rm s}$ is the quantity proportional to the superfluid density $n_{\rm s}$.  Compared with the results in LFAO-F [figure~\ref{G_TFmulti}(d)] and in CFCAF [figure~\ref{G_TFmulti2}(a)], it is noticeable that $\sigma_{\rm s}$ rises relatively sharply just below $T_c$, and becomes mostly independent of temperature below 15 K ($\simeq0.4T_c$).  A curve fit using the power law, $\sigma_{\rm s} =
\sigma_0 [1-(T/T_{\rm c})^\beta]$, yields $\beta= 4.08(5)$, which is perfectly in line with the prediction of conventional BCS model for $s$-wave pairing.  The gap parameter is obtained by a fit using the weak coupling BCS model to yield $\Delta=8.35(6)$ meV and corresponding ratio 
$2\Delta/k_BT_c=5.09(4)$.  These values are consistent with the superconducting order parameter of the isotropic $s$-wave pairing with a relatively strong coupling to some bosonic excitations.

It is inferred from recent angle-resolved photoemission spectroscopy (ARPES) study on BKFA with the same potassium content that the magnitude of superconducting gap depends on the Fermi surfaces \cite{Ding:08}.  They report $\Delta_1\sim12$ meV on small Fermi surfaces and $\Delta_2\sim6$ meV on the large one.  We examined the consistency of our data with the ARPES result by employing the double-gap model shown by eq.~(\ref{dblgap}).
The curve fit assuming a common $T_c=38$ K and a large gap fixed to 12 meV ($2\Delta_1/k_BT_c=7.3$) perfectly reproduces data in figure~\ref{bkfa-chi}(b) with $\Delta_2=6.8(3)$ meV [$2\Delta_2/k_BT_c=4.1(2)$] and the relative weight $w=0.30(3)$, where the obtained curve is virtually identical with that for the single gap on figure~\ref{bkfa-chi}(b).  This, while endorsing the credibility of our data in terms of temperature dependence of $\sigma_s$, indicates that the quasiparticle excitation spectrum associated with multi-gap superconductivity tends to be merged into that of the single gap when the small gap has a large value for $2\Delta/k_BT_c$.

\section{Discussion}

$\mu$SR studies of LFAO-F have been made by a number of groups.
According to those preliminary studies, no clear sign of magnetism is
observed in the sample around $x=0.06$, except for a
weak relaxation observed far below $T_{\rm c}$ for $x=0.05$ and 0.075 in
ZF-$\mu$SR spectra and an unidentified additional relaxation observed in
TF-$\mu$SR spectra for $x=0.075$.\cite{Carlo:08,Luetkens:08-2,Luetkens:08} This led us to
recall the sensitivity to chemical stoichiometry in the emergence of the
spin glass-like $A$-phase observed near the boundary between the
antiferromagnetic and superconducting phases in
CeCu$_2$Si$_2$.\cite{R_CeCu2Si2}.  In addition to the $A$-phase, the
present LFAO-F system exhibits a closer similarity to this classical
heavy-fermion superconductor such as the phase diagram against
pressure/doping.\cite{R_Monthoux} Further study of the dependence of
fractional yield for the magnetic phase with varying $x$ (in small steps
near the phase boundary) would provide more insight into the true nature
of these phases and the mechanism of superconductivity itself that is
working behind the coexistence/competition.

The volumetric expansion of superconducting domains upon electron doping 
to the Fe$_2$As$_2$ layers by Co substitution for Fe is a remarkable feature that has no counterpart in high $T_c$ cuprates.  Meanwhile, this behavior to some extent reminds a parallelism observed in the effect of Zn substitution for Cu in the cuprates that {\sl destroys} superconductivity: it appears that superconductivity is suppressed over a certain domain around the Zn atoms like a ``Swiss cheese" when an extra $d$ electron is introduced\cite{Uemura:03}.  Despite that the effect discussed in the cuprates are completely opposite to that in iron pnictides, the observed ``local" character of doping in CFCAF, which seems to come from a short coherence length $\xi$ (that probably determines the domain size, so that $\xi\sim d_s/2\simeq0.45$ nm), may provide a hint for the microscopic understanding of superconductivity on the Fe$_2$As$_2$ layers, particularly for the $n$ type doping.

In the meantime, the superconducting character of $p$ type iron pnictides seems to be considerably different from $n$ type ones (at least for a doping range near the boundary between magnetic and superconducting phases), as suggested from the behavior of superfluid density
observed by \msr\ in \bkfa. Although the temperature dependence of $\sigma_{\rm s}$ in these compounds can be reproduced by the above mentioned double-gap model, the gap parameters shown in tables \ref{T_SgmPara} and \ref{T_SgmPara2} are considerably smaller than those in BKFA. The double-gap feature revealed by ARPES supports a view that  superconductivity occurs on complex Fermi surfaces consisting of many bands (at least five Fe $d$ bands) that would give rise to certain intricacy \cite{Mazin:08,Kuroki:08}. Apart from the validity of applying the double-gap model to electron-doped iron pnictides, these figures suggest that the hole doping may occur in the bands different from those for electron doping, where the characteristic energy of the Cooper pairing may differ among those bands.   Concerning the competition/coexistense of superconductivity and magnetism, there are so few data available at this stage and thus \msr\ study must be extended over the region of $x$ near the phase boundary (centered around $p=x/2=0.1$).  Here, the control over the homogeneity of the specimen would be a key to proper understanding of the electronic ground state. 

Finally, it would be appropriate to comment on the possible influences of impurities to the temperature dependence of $\sigma_{\rm s}$.  First of all, they serve as the source of scattering for the Cooper pairs, where, depending on the corresponding length of mean free path, the anisotropy of the pair potential tends to be smeared out when $\xi\ge l$.  This effect seemed to be so strong in certain high-$T_c$ cuprates that $\sigma_{\rm s}$ might have shown a behavior resembling that of $s$-wave symmetry\cite{Harshman:87},  leading to some concern that a similar situation might have occurred in \bkfa. 
However, as mentioned earlier, our estimation on $l$ indicates that the condition $\xi\ll l$ is satisfied in our sample,  and therefore we can interpret the temperature dependence of $\sigma_{\rm s}$ without such ambiguity. 

Another important influence of impurities might be that they serve as pinning centers for 
flux lines.  The flux pinning distorts the flux line lattice, leading to additional depolarization to enhance $\sigma_{\rm s}$.  Since the pinning tends to be stronger at lower temperatures, it may distort $\sigma_{\rm s}$ to exhibit stronger temperature dependence
observed for the order parameters with nodes. (Note that it has an effect 
opposite to what is expected for the above-mentioned ``scattering".)  This has been
tested for the case of \cfcaf\ with $x=0.075$ at 2 K by shifting the external field after 
field cooling.  As a result, we found that the average field in the sample indeed followed
the shift of the external field.  Thus, we confirmed that pinning was negligible at least 
in this particular case, and can conclude at this stage that the effect of impurities 
did not affect the temperature dependence of $\sigma_{\rm s}$ seriously in \cfcaf.

\section{Summary}

It has been revealed by our $\mu$SR experiment that superconducting and magnetic phases coexist in \lfaof\ with $x=0.06$. These two phases simultaneously develop just below $T_{\rm c}$, strongly suggesting an intimate and intrinsic relationship between these two phases.  The
result of TF-$\mu$SR measurement suggests that the superconductivity of
LaFeAs(O$_{0.94}$F$_{0.06}$) cannot be explained by the conventional
weak-BCS model (single gap, $s$-wave).  While a similar situation is suggested 
for the superconducting portion of \cfcaf\ with $x=0.075$ and 0.15, the  
segregation between those two phases are more complete, as inferred from the progressive  development of superconducting volume fraction with increasing $x$. This ``local" character seems to be an important feature common to iron pnictides (i.e., an intrinsic property of Fe$_2$As$_2$ layers), probably controlled by the short coherence length ($\simeq0.45$ nm in CFCAF).

Meanwhile, we have shown in a hole-doped iron pnictide, \bkfa, that the superconducting 
order parameter is characterized by a strong coupling to paring bosons, as inferred from large gap parameters ($2\Delta_i/k_BT_c\gg3.53$). The temperature dependence of the superfluid density, $n_{\rm s}(T)$,  determined by \msr\ is perfectly in line with that predicted by the conventional BCS model with fully gapped $s$-wave pairing.  A detailed analysis with phenomenological double-gap model indicates that $n_{\rm s}(T)$ is also consistent with the presence of double-gap, although the large gap parameters make it difficult to determine the multitude of energy gap solely from $n_{\rm s}(T)$.

\ack
The \msr\ studies on iron pnictides cited in this review have been conducted under collaboration with M. Hiraishi, M. Miyazaki,  A. Koda (KEK/Sokendai), Y. Kamihara, S. Matsuishi, H. Hosono (Tokyo Inst.~Tech.), H. Okabe, and J. Akimitsu (Aoyama-Gakuin U.). We would like to thank the TRIUMF staff for their technical support during the $\mu$SR experiment. This work was partially supported by the KEK-MSL Inter-University Program for Oversea Muon Facilities and by a
Grant-in-Aid for Creative Scientific Research on Priority Areas from the
Ministry of Education, Culture, Sports, Science and Technology, Japan.

\end{document}